# Water-in-water PEG/DEX/protein microgel emulsions: effect of microgel particle size on the rate of emulsion phase separation

Andrzej Balis,[1] Georgi Gochev,[1,2]* Domenico Truzzolillo,[3] Dawid Lupa,[4] Liliana Szyk-Warszynska,[1] Jan Zawala[1]

[1] *Jerzy Haber Institute of Catalysis and Surface Chemistry, Polish Academy of Sciences, 30-239 Krakow, Poland*

[2] *Institute of Physical Chemistry, Bulgarian Academy of Sciences, 1113 Sofia, Bulgaria*

[3] *Laboratoire Charles Coulomb, UMR 5221, CNRS-Université de Montpellier, F-34095 Montpellier, France*

[4] *Jagiellonian University, Faculty of Physics, Astronomy, and Applied Computer Science, M. Smoluchowski Institute of Physics, 30-348 Kraków, Poland*

* Corresponding author: georgi.gochev@ikifp.edu.pl

**Abstract**

Protein nanoparticles have been proven to be highly effective stabilizers of water-in-water emulsions obtained from a number of different types of aqueous two-phase systems (ATPS). The emulsion stabilizing efficiency of such particles is attributed to their affinity to the water/water interface of relevant ATPS, and emulsion formulations with long-term stability were reported in the recent years. In this study we investigated the macroscopic dynamics of the early-stage time evolution of dextran-in-polyethylene glycol emulsions obtained from a single ATPS and containing β-lactoglobulin microgel particles of various diameters (*ca*. 40–190 nm). The results revealed the existence of a threshold in microgel size above which the water-in-water emulsion is stabilized, and that the process of segregative phase separation is determined by the interplay of droplets coalescence and sedimentation. Efficient droplet coalescence inhibition was found for microgel particles larger than 60 nm. Based on previous literature results, we discuss our coalescence-driven phase separation data in the context of the formation of durable particle layers on the emulsion droplets and the resulting droplet-droplet interactions.

# 1. Introduction

Water-in-water (W/W) emulsions form in thermodynamically equilibrated aqueous two-phase systems (ATPS) (Esquena, 2016, 2023). The latter are systems, in which segregative phase separation occurs spontaneously due to thermodynamic incompatibility of specific solutes such as hydrophilic polymers and/or salts (Frith, 2010; Iqbal et al., 2016; Chao & Shum, 2020; Bayliss & Schmidt, 2023;). The interest to ATPS is seemingly related to their significant potential for applications in liquid/liquid fractionation processes such as extraction, isolation, purification, partitioning and enrichment of biomolecules and cells (Iqbal et al., 2016; Nadar et al., 2017; Jiang et al., 2019; Chao & Shum, 2020; Hao et al., 2022; Bayliss & Schmidt, 2023; Minagawa et al., 2023). However, industrial applications require development of scale-up strategies (Torres-Acosta et al., 2019). In comparison to "conventional" water-oil based emulsions, the all-aqueous nature of W/W emulsions may render them advantageous in a number of relevant biocompatible applications including foods, encapsulation and delivery systems, as well as biomimetic microreactors (Dewey et al., 2014; Esquena, 2016; Nicolai & Murray, 2017; Dickinson, 2019).

Control over the properties and stability of W/W emulsions, as in the case of any dispersed system, requires understanding of the mechanisms of (de)stabilization as deep as on the nanoscale level. In the case of "conventional" liquid/fluid colloidal systems (soft colloids) such as foams and water-oil based emulsions, the main (de)stabilization mechanisms are already identified (drainage, Ostwald ripening, coalescence) and comparatively well-understood, however, some open questions are still waiting for a solution (Langevin, 2023). W/W emulsions exhibit at least two major problems related to the physicochemical properties of the relevant W/W interface – the remarkably low interfacial tension (on the order of µN/m) and the fundamental point of the definition of the W/W interface (Dickinson, 2019; Chao & Shum, 2020). Concerning the latter, the thermodynamic definition of a (soft) liquid/fluid interface is given by a Gibbs dividing surface. In fact, real liquid/gas and liquid/liquid interfaces possess finite widths, determined by an intrinsic transition zone between the adjacent phases, which in turn, is further modulated by roughness contributions (e.g. capillary waves and structural factors from adsorbed molecules). For bare soft interfaces, the physical interfacial width is of the order of several Angstroms (Braslau et al., 1985; Schlossman, 2005) and when adsorbed molecules, either low-molecular-weight surfactants or amphiphilic macromolecules, are present, the interfacial layer thickness becomes of the order of several nanometers (Lu et al., 2000; Gochev et al., 2024). In contrast, understanding water/water interfaces is still challenging . The width of the interfacial region in W/W emulsions is estimated as being an order of magnitude larger, which exceeds the correlation length of the constituent polymer solutions (Scholten et al., 2004; Tromp et al., 2014; Vis et al., 2018). Therefore, caution should be exercised when discussing the boundary, which separates an aqueous droplet and the aqueous continuous phase in W/W emulsions (Dickinson, 2019), and it would be rather appropriate to speak about an "inter-phase boundary zone". Furthermore, in contrast to water-oil or water-air systems, in W/W emulsions such "boundary zone" is constituted by the same common solvent (water) of both incompatible polymer phases and it is presumably polymer-depleted (Tromp et al., 2014; Vis et al., 2018). In this context, traditional molecular emulsifiers are ineffective as stabilizers of W/W emulsions due to their relatively small size and the fact that their adsorption energy is insufficient in comparison to the thermal motion in the system. An intriguing question is the minimum particle size required for efficient stabilization of these peculiar emulsion systems.

So far, the efforts put to achieve efficient stabilization of W/W emulsions followed two main strategies for enhancing droplet stability (Chao & Shum, 2020; Esquena, 2023): 1) interfacial complexation (formation of membranes over the emulsion droplets); and 2) the Pickering effect by the use of particles of different chemical compositions and shapes (Nicolai & Murray, 2017; Dickinson, 2019). In the latter case, the efficiency of various particles to stabilize Pickering W/W emulsions were investigated and for the case of spherical particles, in particular, it's worth recalling solid silica particles (Murray & Phisarnchananan, 2014; Griffith & Daigle, 2018), synthetic-polymer particles (Balakrishnan et al., 2012; Nguyen et al., 2015; Waldmann et al., 2024), as well as protein particles (Nguyen et al., 2013; Gonzalez-Jordan et al., 2016, 2017, 2018; You et al., 2023; Zhou et al., 2024), just to mention a few.

In Pickering W/W emulsions with protein-based particles, the whey protein β-lactoglobulin (BLG) was mainly used (Nguyen et al., 2013; de Freitas et al., 2016; Gonzalez-Jordan et al., 2016, 2017; Zhang et al., 2021; You et al., 2023). The reason for this is perhaps the fact that BLG-based aggregates with diverse morphologies can be easily incubated with good reproducibility (Jung et al., 2008; Schmitt et al., 2009), and subsequently, serving as efficient stabilizers of W/W emulsions (Gonzalez-Jordan et al., 2016). In (Gonzalez-Jordan et al., 2016), the authors monitored the stability and structure of polyethylene glycol/dextran (PEG/DEX) emulsions as affected by the type of BLG aggregates (fibrils, microgels or fractal aggregates) and found that this effect is pH-dependent (for pH 3 and pH 7). Additionally, it was determined that the emulsion droplet size distribution and the stability of PEG/DEX emulsions are sensitive to the size of the BLG nanoparticles used as stabilizers (Nguyen et al., 2013). Further, BLG microgel particles were used in hydroxypropyl methylcellulose/DEX (Zhang et al., 2021) and xyloglucan/amylopectin (de Freitas et al., 2016) emulsions, with both studies demonstrating a strong influence of pH on the emulsion stability. In addition to this, rheological and tribological properties of gelatinized corn starch/κ-carrageenan emulsions stabilized by whey protein microgel particles were also investigated (You et al., 2023). In our opinion, there are still important aspects of the stability of all-aqueous emulsions that still require further clarification and explanation. For instance, the interrelations between the particle size and other physicochemical characteristics of W/W interfaces and the corresponding emulsions are not well understood. Even the exact mechanisms of the stabilization of W/W emulsions have not been unveiled to date. In foams and oil-water based emulsions, the concept of disjoining pressure barriers (mainly electrostatic and steric) against rupture of the thin liquid films formed between foam bubbles or emulsion droplets has been exploited with great success (Exerowa et al. 2018; Tadros 2016, Langevin, 2023). In the case of colloidal particles at water/air and water/oil interfaces, additional phenomena, such as oscillatory forces, bridging, and capillary effects, have also been identified (Denkov et al., 1992; Aveyard et al., 2003; Horozov 2008).

The present work is, in fact, related to some previous studies on PEG/DEX/BLG microgel emulsions (Nguyen et al., 2013; Gonzalez-Jordan et al., 2016, 2017). In those works, the emulsion long-term stability (on the order of hours and days) was examined via measurements of the droplet size and the rate of creaming/sedimentation. In this study, we turn the attention to the early-stage time evolution (on the order of an hour) of PEG/DEX/BLG microgel emulsions by investigating the macroscopic dynamics of segregative phase separation using a dedicated optical home-made setup. The key objective of our experiments is to get more detailed insights on microscopic processes, namely, to quantify the interfacial affinity of the BLG microgel particles of different size and their ability to diffuse and settle within the "boundary zone" of emulsion droplets. i.e., their stabilizing power against early-stage droplet coalescence. The effects of other influencing factors such as pH and the polymer molecular weight were also monitored, but with less attention. Supporting information was obtained from confocal microscopy observations and interfacial tension measurements.

## 2. Materials and methods

### 2.1. Materials

Polyethylene glycol (PEG) with molecular weight $M_w \approx 10$ kDa was obtained from Sigma Aldrich and Dextran (DEX) with $M_w \approx 75$ kDa was a product by Thermo Scientific. Native β-lactoglobulin ($M_w \approx 18.3$ kDa) in the form of lyophilized powder was supplied by Food Process Engineering, TU Munich, Germany; the procedure for its isolation and purification are described in detail in (Toro-Sierra et al., 2013). Hydrochloric acid was purchased in Chempur (Piekary Slaskie, Poland), while sodium hydroxide was purchased in Sigma Aldrich. Fast Green Dye : (ethyl-[4-[[4-[ethyl-[(3-sulfophenyl)methyl]amino]phenyl]-(4-hydroxy-2-sulfophenyl) methylidene]-1-cyclohexa-2,5-dienylidene]-[(3-sulfophenyl)methyl]azanium) was purchased in Ambeed, IL, USA. Cuvettes for separation of emulsions (4.2 ml, (H x W): 45 x 12 mm, PMMA, transparent with four optical sides) were purchased in Sarstedt. All aqueous solutions were prepared using ultrapure water obtained from Milli-Q Ultrapure Water System (Merck Milipore).

### 2.2. Methods

**Preparation of BLG nanoparticles**

BLG nanoparticles were synthesized based on a previously described protocol (Schmitt et al., 2009). Native BLG solutions with concentration of 1 % wt. were prepared in pure water and stirred for 2 hours. The obtained solution was kept in a fridge overnight to ensure full protein hydration; then the solution was filtered through a sterile PES filter (0.22µm). The filtered solution was divided into 9 portions and for each one a given pH was adjusted (pH-range 5.6–6.4 with 0.1 pH-unit steps) by titration with 0.1 M HCl. Each solution was heated to 85°C in a water bath and kept for 20 minutes without stirring. The obtained dispersions of BLG aggregates were then cooled to room temperature in an ice bath (for about 30 minutes). This allowed us to obtain a series of BLG microgels with different particle sizes, which are abbreviated as BLG($d_h$), where $d_h$ in [nm] is the z-averaged hydrodynamic diameter. Further details are given in section 3.

**Determination of particle size**

The size of the BLG microgel particles was determined by Dynamic Light Scattering using Malvern Panalytical, Model: ZSU3305. Measurements were performed using a disposable polystyrene cuvette (Sarsted, Germany). The z-averaged hydrodynamic diameter $d_h$ with standard deviations, polydispersity index (PDI) as well as intensity and number size distributions from autocorrelation function analysis were evaluated. The measurement data are summarized in Table S1 (Supplementary Materials).

**Atomic Force Microscopy imaging of BLG nanoparticles**

Sample preparation: 10 µL of BLG nanoparticle suspension was diluted with 90 µL of 1 mM HCl. A small drop (40 µL) of suspension obtained in this way was deposited at freshly cleaved V1-grade muscovite mica disc (TED PELLA, USA) and left covered for 15 min. After this period, the surface was rinsed three times with 1 mM HCl, two times with ultrapure water and gently dried in the stream of nitrogen. High-resolution topographical AFM images of BLG nanoparticles deposited on mica were obtained using MultiMode 8 AFM with a Nanoscope 8.0 controller (Bruker Nano Surfaces) operating in a Peak Force Tapping mode. The probes used were standard ScanAsyst-Air cantilevers (Bruker) with a nominal resonant frequency 70 kHz, spring constant of 0.4 N/m and tip radius equal 2 nm. Peak Force setpoint, Peak Force frequency and Peak Force amplitude were set to 200–500 pN, 4 kHz and 30–50 nm, respectively. Depending on the sample, the scanning rate was set in the range of 0.7–1.0 Hz . Raw AFM images were processed using Gwyddion software (Nečas & Klapetek, 2012). Typical procedure consisted of polynomial (1$^{st}$ or 2$^{nd}$ order) background subtraction, step line correction and median row alignment. If needed, scars were automatically removed.

**Preparation of emulsions**

Briefly, 1.5 mL of PEG (20% wt.), 1.5 mL DEX (20% wt.) and 0–0.8 ml BLG aggregates (1% wt.) aqueous solutions (each one set to a given and the same pH (2–6)) were mixed and topped up with water to a total volume of 3.8 ml. Such aqueous mixtures were stirred at 300 rpm for 30 minutes in a water bath set to 25° C. It is worth noting that the order of mixing of solutions or the time and intensity of mixing are not expected to influence significantly the structure and properties of the obtained emulsions (Nguyen et al., 2013; Gonzalez-Jordan et al., 2016, Freitas et al., 2016). After stirring, 3.0 ml of the obtained emulsion was immediately transferred using a pipette to a plastic cuvette, and video recording was initiated. After full phase separation, the equilibrated protein-free ATPS had a composition of 12.5 % wt. (PEG-rich phase) and 21.5 % wt. (DEX-rich phase) (hereafter called simply PEG- or DEX- phase), calculated according to (Dumas et al. 2020) from the known masses of components (water and polymer) and the measured volume fractions of the PEG (66 % vol.) and DEX (34 % vol.) phases. The prepared ATPS belongs to the tie-line region of DEX-in-PEG type of emulsions (data not shown here). In this way, a series of DEX-in-PEG emulsions with various concentrations of different BLG($d_h$) particles were prepared and their early-stage macroscopic dynamics was investigated for at least 75 min. All the experiments with such emulsions were performed at room temperature, T=25° C.

**Confocal Laser Scanning Microscopy imaging**

Emulsion samples for confocal imaging were prepared as explained above with the only difference being the addition of 38 µL of 1 % wt. solution of Fast Green (Murphy et al., 2018) after mixing with DEX and equilibrating the mixture for 30 min with constant stirring to facilitate the binding of dye to the protein. The samples under observation were visualized by fluorescent confocal laser scanning microscopy (CLMS) LSM 780 (Carl Zeiss, Germany), using 20x/0.8 Plan-Apochromat objective. HeNe laser with excitation wavelength of 633 nm (5mW) set at 1.8 % was used as an excitation source. Emission was recorded in the spectral range 644–735 nm through beam splitter 488/561/633, and pinhole 40 mm. Optical zoom 1, master gain 889 and digital gain 0.3 settings were applied.

**Interfacial tension**

Interfacial tension γ measurements were performed with a Krüss spinning drop tensiometer (SDT). The temperature was set at 25.0 ±0.5° C and kept constant using an air flow controlled by a Peltier system. Different tests were performed with rotation rates ranging from 800 to 6900 rpm depending on the sample, such that buoyancy effects were negligible compared to centrifugal forces. Rates of rotation were accurate to 1 %. All drops were illuminated by a blue Light Emitting Diode (LED) with a dominant emission wavelength of 469 nm. Spinning drop measurements leverage on the formation of millimetric droplets in an immiscible and denser background fluid. In our case the lighter fluid (PEG phase) was deposited on one Teflon cap of the capillary, the latter being prefilled with the denser fluid (Dextran phase). The capillary was spun at high angular speed ω > 1000 rpm, that was then reduced by steps Δω. This resulted in the formation of isolated drops due to the viscous pinch-off mechanism induced by surface tension (Stone et al., 1986). The radius of the so formed droplets was in the range 0.75–1 mm. In all cases, the droplets were in full contact with the surrounding fluid once the capillary was spun.

**Examination of phase separation dynamics**

A home-made device was developed to obtain quantitative information about the rate of phase separation in the investigated emulsions. The general scheme of the device is presented in Fig. 1. It consists of a FLIR Blackfly BFS-U3-23S3M-C camera and an AF-S Micro Nikkor 60 mm 1:2.8 G lens. The sample was illuminated by a KL 2500 LCD lamp from Fiber Optics Schott AG. Images were captured at a frequency of 2 Hz in automatic exposure mode. The sample was placed in a plastic 3D-printed chamber connected to a Thermo Scientific SC150 immersion circulator with Arctic bath A25 thermostat. To illustrate the measurements procedure, a representative scheme of the analysis of the rate of phase separation in an emulsion sample is presented in Fig. 2. In fact, the velocity $V$ of the advancing front line of phase separation (hereafter called simply *front*) is determined by measuring the displacement $h$ of the upper (PEG phase solution/emulsion) and/or the lower (emulsion/DEX phase solution) phase boundaries (fronts) with time $t$. Raw data in terms of front displacement $h(t)$ profiles (Fig. 2A) were obtained from image analysis of video recordings (spatiotemporal resolution of 2 FPS and 1 pixel ≡ 0.017 mm) using automated scripts written in Python, and then converted into linear velocity $V(t)$ profiles (Fig. 2B).

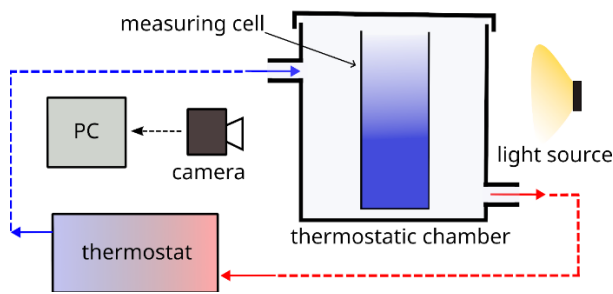

**Fig. 1.** Scheme of the phase separation device.

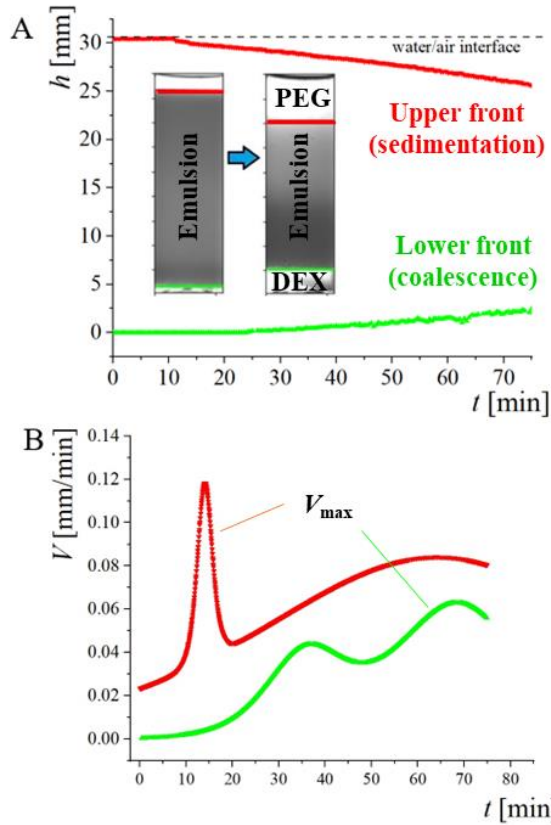

**Fig. 2.** Representative sample of a DEX-in-PEG/BLG(81) emulsion at pH 3. **A.** $h(t)$ profiles for the displacement of the upper and lower horizontal phase boundaries (fronts). **B.** Corresponding $V(t)$ profiles for the time evolution of the velocity of front displacement.

Analysis of the data of the emulsion sample from Fig. 2 suggests that the early-stage time evolution of this emulsion may involve two concurrent processes. Droplet coalescence is a destabilization process that leads to a progressive increase in droplet size, *i.e.*, a decrease of the emulsion dispersity (the kinetics of this process could not be followed in this study). Large enough DEX droplets or flocks of smaller droplets will sediment due to gravity, leaving supernatant macroscopic PEG phase and a concomitant appearance of an *upper front*. Further coalescence will lead to the formation of a subnatant macroscopic DEX phase and a concomitant appearance of a *lower front*. Hence, the displacement of the lower front was used as a measure for the rate of DEX phase separation due to coalescence of DEX droplets, whereas, the displacement of the upper front was used as a measure for the rate of PEG phase separation due to droplets sedimentation. Fig. 2B shows peaks in the $V(t)$ profiles. Increase in the intensity of the velocity peaks (maximum velocity $V_{max}$) and their shift towards shorter times were used as quantitative measures for the rate of phase separation of either of the two polymer phases.

## 3. Results and discussion

### 3.1. BLG nanoparticles characterization

As mentioned above, thermal aggregation of BLG leads to the formation of supramolecular structures of various morphologies and dimensions depending primarily on the solvent conditions (pH; inorganic ions), but also on the concentration of the native protein solution (Jung et al., 2008; Schmitt et al., 2009; Phan-Xuan et al., 2011). Heating of BLG solutions at pH 2 leads to polypeptide hydrolysis and subsequent incubation of amyloid fibrils built up by a certain fraction of amyloidogenic hydrolysates (Akkermans et al., 2008; Loveday et al., 2017). At pH near the isoelectric point (pI ≈ 5.1) (Engelhardt et al., 2013), the formation of BLG particulate aggregates with a relatively narrow size distributions is observed, with sizes between 100 nm and a few microns depending on the thermal incubation history (Donald, 2008). The heating of BLG solutions set to a narrow pH range of 5.8–6.2 at 85° C results in the formation of nearly spherical microgel particles characterized by relatively low polydispersity index (6–17 %, see Table S1 in Supplementary Materials) and hydrodynamic diameters $d_h$ of *ca*. 50–200 nm (Jung et al., 2008; Schmitt et al., 2009; 2010; Phan-Xuan et al., 2011; Zhang et al., 2020, 2021). At higher pH-values (up to pH 7) in salt-free BLG solutions, relatively small (below 50 nm) worm-like primary aggregates are formed, which randomly organize in fractal clusters at high concentrations (Jung et al., 2008; Mehalebi et al., 2008).

The microgel incubation procedure was applied to 1 % wt. native BLG solutions set to a given pH-value in the range 5.6–6.4 with 0.1 pH-unit steps; photographs of the native BLG solutions and the resulting microgel dispersions are shown in Fig. S1 (Supplementary Materials). Microgel dispersions incubated at pH 5.6 and pH 5.7 exhibited clumpy white precipitates, whereas the rest of the native solutions formed stable microgel dispersions. These results are in accordance with the literature (Phan-Xuan et al., 2011).

Supporting experiments with AFM revealed that the BLG microgel particles under examination are spherical and appear mostly as isolated entities. Topographical images, together with size distributions of exemplary BLG microgels (as incubated) are presented in Fig. S2 (Supplementary Materials). The microgel particle size was determined by DLS and the results from the analysis are tabulated in Table S1 (Supplementary Materials), including the z-averaged size PDI, obtained via cumulant analysis of the intermediate scattering function as well as mean values for volume-weighted and number-weighted size distributions (via General Purpose/Multiple Narrow Mode algorithm). Fig. 3 (main panel) presents the z-averaged hydrodynamic diameter $d_h$ as a function of the initial pH of the native BLG solutions. The two shown datasets correspond to measurements of the original microgel dispersions ($d_{h,O}$) and those of the same dispersions after adjustment to pH 3 ($d_{h,S}$). The data for $d_{h,O}$ are in a very good agreement with previous reports (Schmitt et al., 2009; Phan-Xuan et al., 2011). It is evident that the change in pH of the original dispersions to pH 3 causes a systematic swelling of the microgel particles. The degree of swelling was estimated via the simple relation $\frac{d_{h,S} - d_{h,O}}{d_{h,O}}$ and is presented as a function of $d_{h,O}$ in the inset of Fig. 3. Concerning the particles sizes with $d_{h,O} \approx 50\text{–}160$ nm (obtained at pH 5.8–6.2) for the typical BLG microgels, the results reveal swelling of 15–35 %, which is in agreement with literature data (Schmitt et al., 2010; Zhang et al., 2020). However, we observe here that the degree of swelling increases with decreasing particle size. We speculate that this effect is related to restricted penetration of solvent into the particle's body and formation of a swollen, softer and less compact corona over the particle's core, as the latter remains unaltered. Such a scenario would therefore lead to stronger swelling-induced relative changes in size for smaller particles. In the outmost regions of BLG microgel sizes, the structure move to other morphologies, namely towards smaller BLG worm-like primary aggregates ($d_{h,O} \approx 30$ nm, PDI 0.23 and $d_{h,S} \approx 40$ nm, PDI 0.23, correspondingly) incubated at pH 6.4 or towards larger BLG particulate aggregates ($d_{h,O} \approx 700$ nm, PDI 0.33 and $d_{h,S} \approx 1000$ nm, PDI 0.75, correspondingly) incubated at pH 5.7.

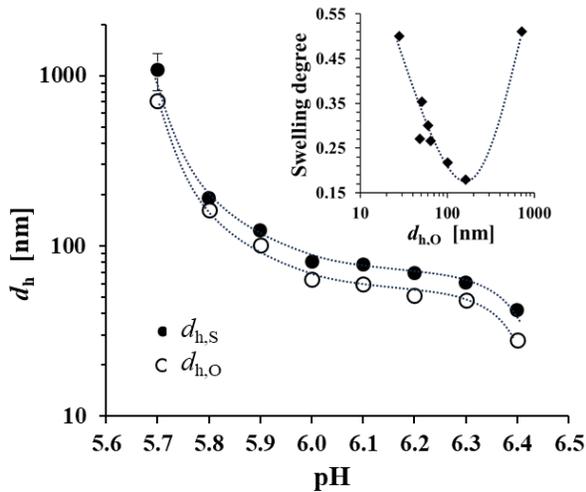

**Fig. 3.** Z-averaged hydrodynamic diameter $d_h$ of BLG aggregates as a function of the pH of the initial native BLG solutions; $d_{h,O}$ – original aggregates dispersions, $d_{h,S}$ – swollen aggregates dispersions after adjustment to pH 3 (error bars for the standard deviations of $d_h$ are smaller than the symbols' size, except for $d_{h,S} \approx$ 1000 nm). Inset: dependency of the degree of aggregates swelling defined as $\frac{d_{h,S} - d_{h,O}}{d_{h,O}}$ on the particle size $d_{h,O}$.

### 3.2. Effect of pH on emulsion stability

The effect of pH on the stability of PEG/DEX emulsions containing BLG aggregates was related to preferential partitioning of such aggregates to the PEG or the DEX phase (Gonzalez-Jordan et al., 2016, 2017; Nguyen et al., 2013). In those studies, the authors demonstrated that all of the three investigated types of BLG aggregates (microgels, fractals or fibrils) partition to the DEX phase at pH 7 (BLG negative net charge) and to the PEG phase at pH 3 (BLG positive net charge). In this study we chose an arbitrary microgel particle size of $d_{h,O} \approx 64$ nm (incubated at pH 6.0). In the following we present results for the rate of phase separation of DEX-in-PEG/BLG(64) emulsions set to several pH values in the range 2–6 (note that $d_{h,O} \approx 64$ nm, incubated at pH 6, is a subject to swelling/compression by the different pH settings, e.g. $d_{h,S} \approx 81$ nm for swollen particles at pH 3).

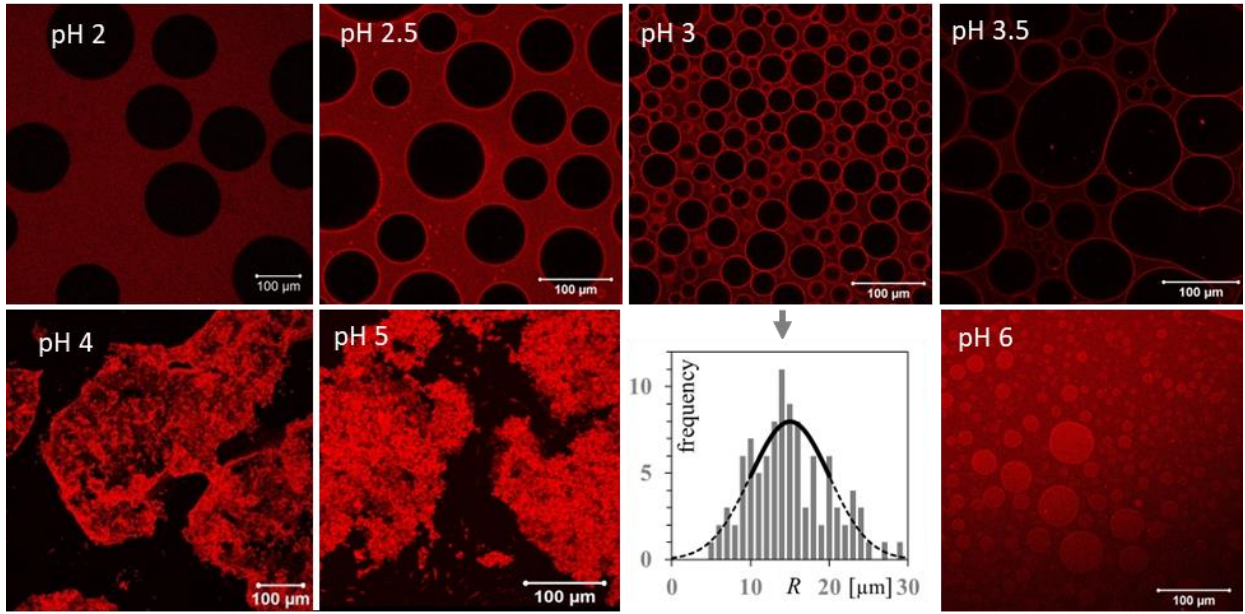

**Fig. 4.** Representative Confocal Laser Scanning Microscopy images of DEX-in-PEG/BLG particles ($d_{h,O} \approx$ 64 nm; at pH 3 $d_{h,S} \approx 81$ nm) emulsions (0.2 % wt. BLG) at various pH, aged for about 30 min. The inset graph presents the relative frequency of appearance of droplets with radius $R$ obtained from manual image analysis (via ImageJ) of 100 droplets in the photograph of the emulsion at pH 3; the solid line is a normal distribution curve centered at 15 µm with a standard deviation of ±5 µm.

Fig. 4 presents Confocal Laser Scanning Microscopy images of such emulsions aged for about 30 min. For the emulsions at pH 2–3.5, the signal from the Fast Green/BLG particle complexes is entirely within the continuous PEG phase. At pH 2, it can be observed that the complexes are not accumulating at the boundaries of the DEX droplets, as indicated by the absence of a halo. On the contrary, accumulation of the complexes at the boundaries of the DEX droplets is unambiguously evident in emulsions at pH 2.5–3.5. Nevertheless, at pH ≥ 3.5 some deformations in the spherical shape of the DEX droplets are observable. For pH 4 and pH 5 (close to the BLG isoelectric point of pI ≈ 5.1) clustering of BLG particles and formation of large non-spherical DEX domains was observed. Such a behavior is analogous to that observed in similar DEX-in-PEG/BLG microgel emulsions (Gonzalez-Jordan et al., 2017) and in emulsions based on hydroxypropyl methylcellulose/DEX ATPS (Zhang et al., 2021) or xyloglucan/amylopectin ATPS (de Freitas et al., 2016). It is noteworthy that at pH 6 the droplets remain visible, yet the affinity of BLG particles to the dispersed DEX phase is more pronounced than in the rest of cases. This is demonstrated by the uniform distribution of BLG particles across both phases. The absence of a substantial enhancement of the signal at the droplet boundaries suggests a weak droplet stabilization.

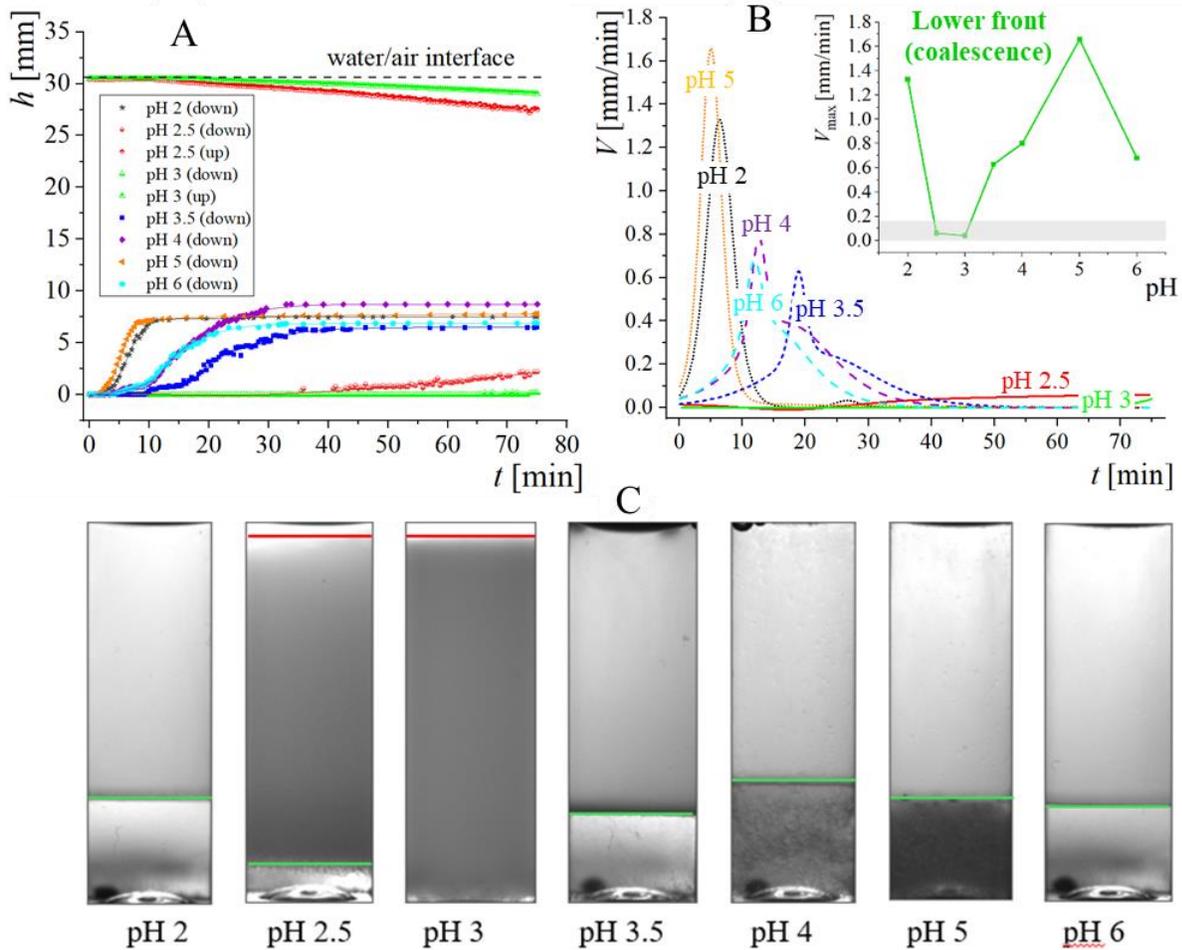

**Fig. 5.** Front displacement experiments for DEX-in-PEG/BLG(64) emulsions (0.1 % wt. BLG) at various pH values. **A.** $h(t)$ plots ("up" means upper front, "down" means lower front). **B.** Corresponding $V(t)$ plots for the lower front. Inset in (B): the evaluated maximum velocity $V_{max}$ of front displacement as a function of pH; the grey-shaded area of up to 0.2 mm/min highlights the lowest rates of coalescence for the most stable emulsions. **C.** Representative snapshots of the emulsions after 75 min from formation, taken from Movie 1 (Supplementary Materials).

Further, the DEX-in-PEG/BLG(64) emulsions were investigated by phase separation tests and Fig. 5A presents the measured $h(t)$ plots. The data for the lower front are transformed into $V(t)$ profiles in Fig. 5B, where the intensity of the velocity peaks (maximum velocity $V_{max}$) and their shift towards shorter times are used as quantitative measures for the increasing rate of DEX phase separation due to DEX droplets coalescence. The data clearly indicate that the lowest rates of DEX phase separation, *i.e.*, the highest emulsion stability, are achieved at pH 2.5 and pH 3 and corroborate the confocal microscopy observations, where the stained BLG particles exhibit the highest concentrations at the emulsion droplets boundaries. The highest rates of DEX phase separation were found for emulsions at pH 2 and pH 5, which are the least stable ones. At pH 5, the BLG particles are very close to pI, which promotes aggregation (visible with naked eye in Fig. 5C) and thus rapid droplets coalescence. At pH 2, the rapid coalescence occurs presumably due to screening of the droplet surface charge by the high ionic strength (Zhang et al., 2021). A summary of these results is presented in the inset of Fig. 5B in terms of $V_{max}$ of the lower front as a function of pH. Note that the maximum velocity values for the emulsions at pH 2.5 and pH 3 are as low as $V_{max} < 0.1$ mm/min. Fig. 5C, supported by Movie 1 (Supplementary Materials), shows the state of the system after 75 minutes. All these clearly indicate that despite the relatively low concentration of BLG nanoparticles (0.1%), the systems at the optimal pH 2.5 and pH 3 are the most stable, which is in agreement with literature reports (Gonzalez-Jordan et al., 2016). For the emulsion at pH 3, after an ageing time of about 30 min, we have estimated the number-average droplet radius $R$ distribution ($R = 15 \pm 5$ µm) by manual image analysis (open-source software ImageJ) of 100 droplets in the respective photograph in Fig. 4. It should be noted that a comparison with literature data is not straightforward, because of differences in the emulsion compositions, such as microgel particle size, polymer $M_w$ and concentrations, phase volume fractions, and other factors, which were found to affect the emulsion behavior and stability in previous studies (Nguyen et al., 2013; Gonzalez-Jordan et al., 2016). For example, $R = 4.5 \pm 2.5$ µm was found for "just-prepared" DEX(160 kDa)-in-PEG(200 kDa) emulsions (25 % vol. of DEX droplets, tie-line interfacial tension of 75 µN/m) with BLG microgels with $d_h \approx 300$ nm at pH 3; this droplet size was found weakly dependent on the protein concentration (0.05–0.3 % wt.) and undergoing insignificant changes with aging of the emulsion (Gonzalez-Jordan et al., 2016). In comparison, our results for the 30 min aged emulsion show a nearly 3-fold larger $R$-values.

### 3.3. Effect of the particle size on emulsion stability

The next and the main step in this study was to examine the effect of the size of BLG microgel particle on the rate of emulsion phase separation. For this purpose we used DEX-in-PEG/BLG($d_{h,S}$) emulsions at 0.1 % wt. BLG and pH 3. The results for the different particle sizes $d_{h,S}$ are presented in Fig. 6 and supported by Movie 2 (Supplementary Materials).

It should be noted firstly that complete phase separation for the emulsion system prepared in the absence of BLG particles occurred after about 10 min. According to the present results, the phase separation behavior of the emulsions containing BLG particles can be divided into three groups according to the particle size $d_{h,S}$ of the protein microgels:

(I) Smaller-sized particles BLG(42) and BLG(61): Only the formation of a lower front (emulsion/DEX phase) was detected. For the smallest microgel particles BLG(42) the rate of DEX phase separation is distinctly faster (high $V_{max}$ peak at about 20 min) as compared to that observed for the emulsion with BLG(61) particles (lower $V_{max}$ peak at about 40 min). These two emulsions are rather unstable, especially the BLG(42) one. It should be noted that in these two experiments we could not clearly detect an upper front. A careful inspection of Movie 2 (Supplementary Materials) suggests that the top liquid fraction is a mixture of PEG solution and rapidly coalescing DEX domains, presumably undergoing spinodal decomposition. Notably, the protein free emulsion showed a very similar behavior, but the top liquid fraction started to turn clearer (PEG phase separation) at earlier stages.

(II) Medium-sized particles BLG(69), BLG(78) and BLG(81): two fronts were detected. The corresponding front velocity profiles $V(t)$ are very similar revealing that the processes of coalescence and sedimentation proceed in parallel, but at relatively slow rates. In these emulsions, the DEX droplets are presumably smaller than those for the emulsions from group (III) and therefore the PEG phase separation is slower due to weak emulsion sedimentation. This explanation is in accordance with literature results, which demonstrated that decrease of the BLG microgel particle size leads to decrease of the droplet size $R$ in PEG-in-DEX emulsions (Nguyen et al., 2013). On the other hand, the appearance and the comparatively slow ascend of the lower front suggest that these emulsions exhibit somewhat lower (but still substantial) stability than those from group (III). The highest fraction of *ca.* 90 % vol. of intact emulsion remaining after 75 min was detected for the BLG(78) sample.

(III) Larger-sized particles BLG(123) and BLG(191): only the formation of an upper front (PEG phase/emulsion) was detected. For the BLG(191) emulsion, a high peak ($V_{max}$) in the $V(t)$ profile appeared at about 15 min and then the velocity levelled off to a very low value at about 50 min. The lack of a lower front and the observed comparatively fast PEG phase separation suggest that this emulsion is stable, but presumably consists of relatively large droplets (or flocks), which cause sedimentation of the emulsion (Nguyen et al., 2013). Similar behavior was observed for the case of BLG(123) particles – the emulsion is also stable, but here the rate of its sedimentation is lower, presumably due to a decrease of the droplets (or flocks) size.

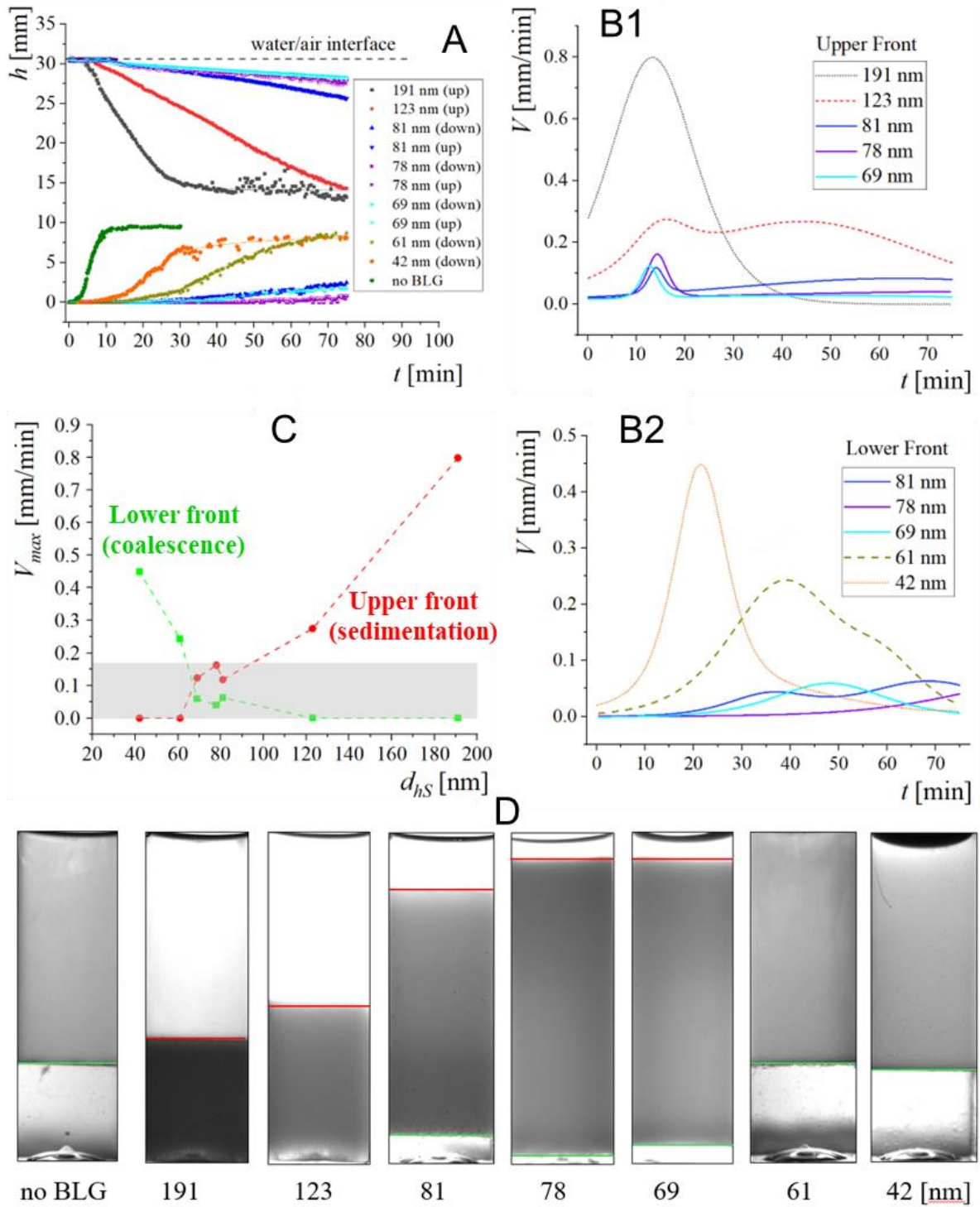

**Fig. 6.** Front displacement experiments for DEX-in-PEG emulsions with various particle sizes $d_{h,S}$ (0.1 % wt. BLG) at pH 3. **A.** $h(t)$ profiles. **B.** Corresponding $V(t)$ profiles for the upper (B1) and the lower (B2) fronts. **C.** The maximum velocity $V_{max}$ of front displacement as a function of $d_{h,S}$; the grey-shaded area of up to 0.2 mm/min highlights the intermediate rates of coalescence and sedimentation. **D.** Representative images of the emulsions after 75 min from formation (note a snapshot for a protein-free blank sample is taken after 30 min).

To further the investigation, we performed interfacial tension γ measurements in the presence of BLG microgels to test their affinity to the interface. The interfacial tension of the pure equilibrated ATPS at pH 3 was firstly measured as a reference: we obtained $\gamma_0 = 67 \pm 1$ µN/m. The magnitude of $\gamma_0$ can be related to the length (TLL) of the tie-line (by which a given sample is characterized within the respective ATPS phase diagram) by several simple relations (Forciniti et al., 1990). However, in the recent research, a power law relation: $\gamma_0 \propto \text{TLL}^p$ was employed either for PEG(200 kDa)/DEX(500 kDa) ATPS with $p = 3.9$ (Balakrishnan et al., 2012) or for xyloglucan(13 MDa)/amylopectin(2.9 MDa) ATPS with $p = 5.7$ (de Freitas et al., 2016). Notably, it was demonstrated that either TLL or $p$ are dependent on the polymer $M_w$ (Forciniti et al., 1990, 1991). At this stage, we haven't measured the TLL of the tie-line that characterizes the particular ATPS used in this study (12.5 % wt. PEG(10 kDa) / 21.5 % wt. DEX(75 kDa)), and we are not able to accurately calculate it on the basis of $\gamma_0$ without unambiguously knowing an actual $p$-value.

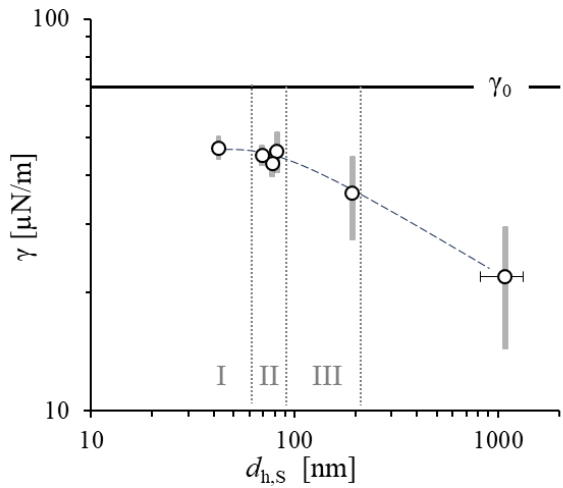

**Fig. 7.** PEG/DEX/BLG($d_{h,S}$) ATPS at 0.1 % wt. BLG and pH 3. Interfacial tension γ of a spinning PEG drop in a DEX matrix as a function of the microgel size $d_{h,S}$; the interfacial tension ($\gamma_0 = 67 \pm 1$ µN/m) of the protein-free ATPS is indicated by a horizontal solid line; the dotted line through the symbols is guiding the eye. The vertical grey-shaded ribbons through the symbols are not standard deviations from measurements, but they reflect the dispersion of the interfacial tension hysteresis shown in Fig. 8 and discussed in the text; the horizontal error bars are standard deviations of $d_{h,S}$. The labels I–III indicate the division of the tested emulsions into three groups according to their phase separation behavior, as explained in the text.

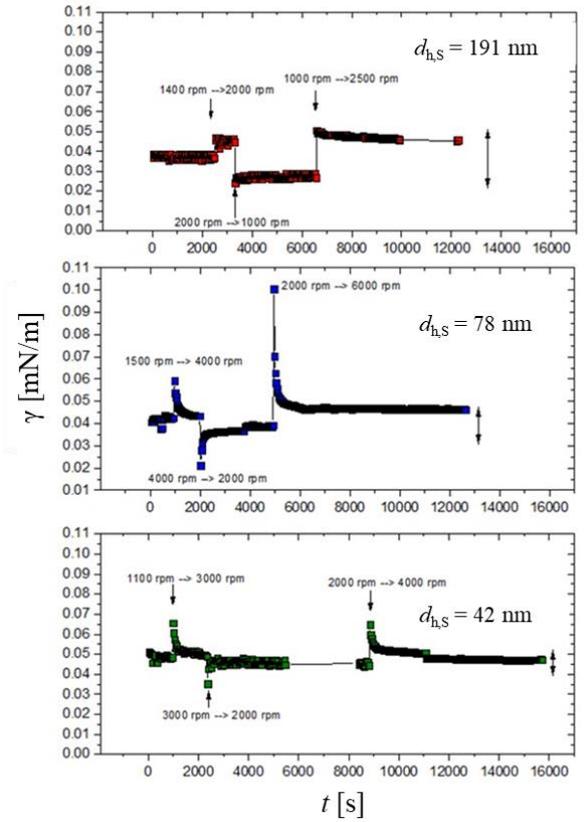

**Fig. 8.** Relaxation time experiments with spinning drops for three different BLG particle sizes $d_{h,S}$ as denoted on the graphs. Arrows point at a sudden change in rotation speed $w_{initial}$(rpm)--> $w_{final}$(rpm). The vertical double arrows are only a guide for the eye and reflect the dispersion of the quasi-equilibrium interfacial tension accessible to our experiments.

The results in Fig. 7 reveal that the addition of a fixed BLG microgel concentration of 0.1 % wt. to the ATPS used leads to a distinct reduction of the interfacial tension $\Delta\gamma = \gamma_0 - \gamma$, which can be interpreted as surface pressure exerted by adsorbed particles. For smaller microgel particles (*ca.* $d_{h,S} \approx 40-80$ nm), the values for $\Delta\gamma$ of about 20 µN/m are virtually independent of the particle size. Increase of the particle size for BLG(191) resulted in increase of $\Delta\gamma$ to 31 µN/m and further to $\Delta\gamma$ = 45 µN/m for the larger polydisperse particulate aggregates (incubated at pH 5.7). For comparison, the interfacial tension $\gamma_0$ = 70 µN/m of a particular PEG/DEX ATPS, measured with the spinning drop method, was reported to decrease with $\Delta\gamma$ = 10 µN/m in the presence of ~200 nm β-casein particles at 0.1 % wt. (Zhou et al., 2024). Our experiments revealed hysteresis of the interfacial tension response to changes in the angular speed ω of the spinning drop. This is a consequence of some peculiarities observed in the spinning drop experiments. During several hours of examination, ω was changed several times (increased or decreased) while γ was measured every 30 s. The results obtained for three different particle sizes $d_{h,S}$, namely the microgel particles BLG(191) and BLG(78), and the worm-like aggregates BLG(42), are presented in Fig. 8. For the BLG(191) microgel we observed a drastic (almost two-fold) change of the "quasi-equilibrium" interfacial tension with variations of ω (denoted on the graphs in Fig. 8). The higher ω (the more stretched the droplet) the higher γ. Due to the sudden increase or decrease of the droplet area and thus of the particles concentration at the interface, the droplets are stretched or compressed, and the particles exchange between the bulk and the interface is slow. This phenomenon is sensitive to the particle diffusivity, which is apparently much slower than the molecular rearrangements of the bare liquid-liquid interface, and consequently, the adsorption kinetics is slower. As a result, the interface gets stuck in a "long-living" metastable state (Keal et al. 2018). For a system containing larger BLG particles, it is therefore more troublesome to determine the precise γ-value, and as a result, its dispersion is relatively high as shown in Fig. 7. On the other hand, for the smallest particles BLG(42) (worm-like aggregates), the effect of ω is much weaker: the interfacial tension relaxes always to the same value within the time accessible to our experiments. Therefore, the smaller the particle size, the faster the system relaxes towards its equilibrated state. The γ(ω) semi-dispersions, indicated in Fig. 7, are as high as about 35 % for the large polydisperse particulate aggregates (incubated at pH 5.7) and decrease to 7–12 % for the microgel particles. Notably, such dynamic hysteresis was not observed for the protein-free ATPS (data not shown) and the above-mentioned value $\gamma_0 = 67 \pm 1$ µN/m was measured with a high accuracy.

In this context, in the particular PEG/DEX system investigated here, at pH 3 the BLG particles should be allowed to migrate from the continuous PEG phase towards the "boundary zone", where the particles get confined, because of their poor ability to partition into the DEX droplet phase (Gonzalez-Jordan et al., 2016), evidenced from the confocal microscopy observations in Fig. 4. The adsorption energy $-\Delta G$ of a spherical particle attached to an interface, taken equal to the free energy $\Delta G$ needed to detach the same particle from the interface (Aveyard et al., 2003), can be estimated as:

$$-\Delta G = \pi r^2 \gamma_0 (1 - |\cos\theta|)^2 \qquad (1)$$

where $r$ is the particle's radius and $\theta$ is the so called Young contact angle, defined as $|\cos\theta| = (\gamma_{P1} - \gamma_{P2})/\gamma_0$ ($\gamma_{P1}$ and $\gamma_{P2}$ being the surface tensions of the particle with each of the polymer solutions). Employing Eq. (1) with $\gamma_0 = 67 \pm 1$ µN/m and approximating $r$ to $d_{h,S}/2$, the calculated adsorption energies, normalized by the thermal energy, i.e. $-\Delta \bar{G} \equiv -\Delta G/kT$, are plotted in Fig. 9A as a function of $|\cos\theta|$ for the different BLG microgel sizes $d_{h,S}$. The magnitude of the adsorption energy may vary from the level of the thermal energy ($-\Delta \bar{G} = 1$) for $\theta \approx 150°–160°$ depending on the particle sizes investigated here (Fig. 9C), i.e., when a particle is faintly attached to the interface, up to about $-\Delta \bar{G} \approx 470$ for the extreme case when the largest particles studied here are optimally pinned to the interface ideally at $\theta = 90°$ (Fig. 9B). Notably, the contact angle can be tuned in a wide range by varying the affinity of the particles to each of the polymer phases (Xue et

al. 2017; Meng & Nicolai, 2023; Waldmann et al., 2024). For emulsion systems where the particles preferentially partition to one of the polymer phases values of |cosθ| > 0.6 were found (Balakrishnan et al., 2012; Keal et al. 2018; Zhou et al., 2024).

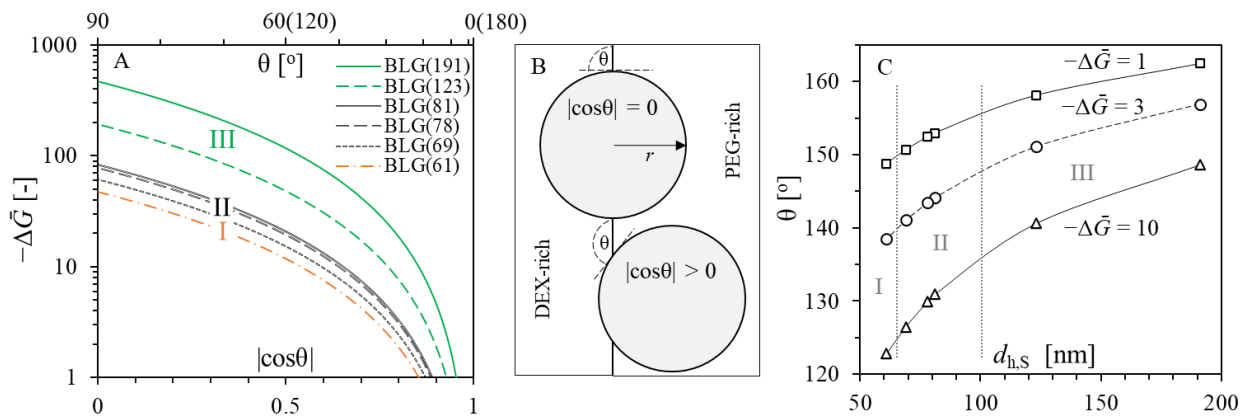

**Fig. 9. A.** PEG/DEX/BLG($d_{h,S}$) ATPS at 0.1 % wt. BLG and pH 3. Normalized adsorption energy $-\Delta\bar{G} \equiv -\Delta G/kT$ ($kT \approx 4.12 \times 10^{-21}$ J at $T = 25°$ C) of spherical microgel particles of different sizes $r = d_{h,S}/2$ as a function of the contact angle θ, calculated via Eq. (1) with $\gamma_0 = 67$ µN/m. **B.** Schematic representation (not to scale) of identical BLG microgel particles with radius $r$, migrated from the continuous PEG phase and adsorbed at the surface of a DEX droplet at different contact angles. **C.** Variation of the contact angle θ with the microgel particles size $d_{h,S}$ for three arbitrary adsorption energies $-\Delta\bar{G}$ of 1, 3 and 10 (lines are guides to the eye). The labels I–III in (A) and (C) indicate the division of the tested emulsions (see Fig. 6) into three groups according to their phase separation behavior as explained in the main text.

We have therefore estimated the adsorption energies of the of BLG microgel particles at the surface of DEX droplets for the emulsion systems studied here (see Fig. 6). The major obstacle that hinders correct analysis is the problem with the evaluation of θ. In situ θ-measurements have been achieved using Confocal Laser Scanning Microscopy for spherical latex particles adsorbed at PEG/DEX (Balakrishnan et al., 2012) or DEX/gelatin (Keal et al. 2018) interfaces, yielding |cosθ| ≈ 0.77–0.87 and |cosθ| ≈ 0.93–0.97, respectively. In those studies, the particle sizes span from a couple of hundred nanometers to several microns and such measurements do not seem feasible for the case of smaller particles (Nguyen et al., 2013; Gonzalez-Jordan et al., 2017). Therefore, we could only indirectly evaluate realistic adsorption energies of the BLG microgel particles investigated.

In the following analysis we exclude the case of the BLG(42) particles, because, as mentioned above, they were found to have different morphology than a spherical shape (Jung et al., 2008; Mehalebi et al., 2008). Furthermore, the high rate of coalescence-driven phase separation measured for the BLG(42) emulsion is in agreement with results for similarly-sized BLG microgel particles in PEG-in-DEX emulsions at pH 7 (Nguyen et al., 2013) as well as for particles from other proteins, namely bovine serum albumin particles (≈50 nm) or ovalbumin particles (≈10 nm) in DEX-in-PEG emulsions (Zhou et al., 2024). Whatever the physical explanation, all these observations corroborate the concept that particles with sizes comparable to the width of the W/W "boundary zone" do not exhibit stabilizing efficiency in W/W emulsions.

If the reason for the comparatively very low stability of the emulsion in the presence of the smallest microgels, namely BLG(61) from group I, was the lack of sufficient particle adsorption at the DEX droplets, ($-\Delta \bar{G} = 1-3$) these particles would have contact angles of about 138º–148º as illustrated in Fig. 9C. Yet, if the BLG(61) particles were still able to be adsorbed at a certain surface coverage on the DEX droplets, the adsorption layer would be seemingly loose and not able to prevent droplets coalescence.

The rate of coalescence-driven phase separation significantly decreased in the emulsions from group II ($d_{h,S} \approx 70-80$ nm) and was practically arrested at metastable states at the end of the experimental time window of 75 min for the emulsions from group III ($d_{h,S} > 120$ nm). The increased stability of these emulsions should be attributed to effective inhibition of droplets coalescence due to the formation of durable interfacial layers of particles, which are resistant to eventual shear and dilational disturbances. Just for illustration, an adsorption energy of $-\Delta \bar{G} = 10$ yields contact angles of about 126º–131º (group II) and 140º–148º (group III) (Fig. 9C). The main feature for the emulsions from these two groups is that the rate of sedimentation increases with the particle size, which should be attributed to an increase of the droplet size due to the decrease of the droplet surface coverage per unit area in the case of larger particles as compared to smaller particles at a given total protein concentration (Nguyen et al., 2013).

The above analysis, based on evaluation of the adsorption energy of the BLG particles is helpful to get insights into the possible structure of the particle's interfacial layers on the droplet surface. Notably, for physically correct quantification of the adsorption energies of the studied nanometric particles and the relevant contact angles, the modified Young's law should be used. The latter accounts for the contribution of a line tension $\tau$, which has been evaluated from theoretical and experimental considerations to span over several orders of magnitude, *ca.* $\tau = 10^{-12}-10^{-5}$ N (Bresme & Quirke 1999; Amirfazli & Neumann 2004, McBride & Law 2012). For the case of particles adsorbed at water/air interfaces, significant effects of $\tau$ on $\theta$ is expected for particle radii below *ca.* 500 nm and when $|\cos\theta| > 0.5$ (Zeng et al. 2011; McBride & Law 2012). Furthermore, comparative studies for liquid/fluid interfaces with different surface tensions $\gamma_0$ revealed that the effect of the line tension on the Young contact angle of particles becomes stronger with decrease of $\gamma_0$ (Bresme & Quirke 1999). Therefore, a correction accounting for $\tau$ will lead to quantitative modifications of the data in Fig. 9A,C, without changing qualitatively our conclusions.

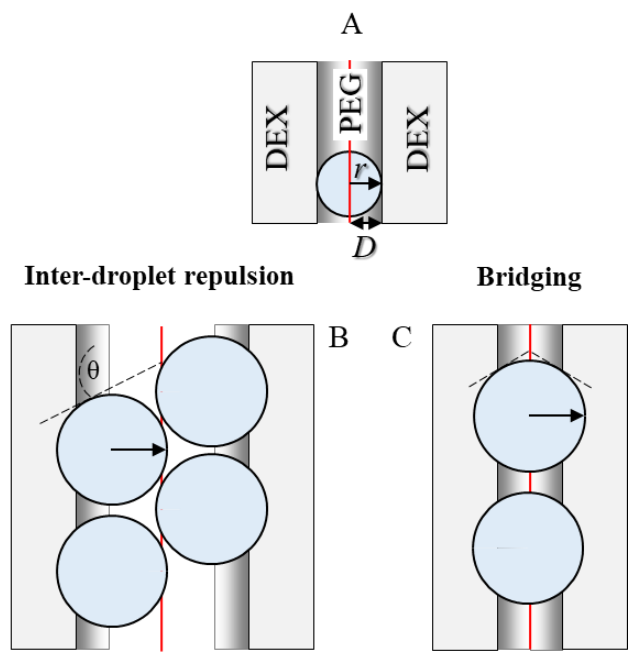

**Fig. 10.** Interaction of two DEX droplets in continuous PEG phase containing preferentially partitioned spherical particles with different radii $r$. $D$ is the interfacial width and $\theta > 90°$ is the contact angle.

The polymer-depleted "boundary zone" can be considered as an extra layer of liquid water with a certain width $D$ separating an emulsion droplet from the continuous phase (Fig. 10) (Tromp et al., 2014; Vis et al., 2018; Dickinson 2019). Such a layer of elevated water content in an ATPS implies preferential hydration and partitioning of hydrophilic solutes other than the pair of incompatible polymers. Such a consideration sheds light on the driving force for particles migrating from the continuous phase to settle within the "boundary zone". It was commented in the literature that a particle with a size comparable to the "boundary zone" width, say $r \approx D$, cannot induce significant interfacial deformation and do not undergo pinning as illustrated in Fig. 10A. Hence, effective particle pinning at the droplet's surface would occur at a certain particle size limit $r/D$. In the present study, effective droplet coalescence inhibition by BLG microgel particles was detected already at about $d_h/D \approx 4$–$8$ for the particles from group II, taking $D \approx 10$–$20$ nm. The latter value was evaluated from theoretical analysis based on scaling approaches for semi-dilute polymer solutions; note $D$ is dependent on the macromolecular structural characteristics (mainly the degree of polymerization and the "blob" size) of the polymers used, as well as on their concentration (Tromp et al., 2014, Vis et al., 2018).

Say, the particles adsorption energy is sufficiently high and they adsorb on the emulsion droplets, yet this may not be the sufficient condition for inhibiting droplets coalescence (Balakrishnan et al., 2012; Gonzalez-Jordan et al., 2018). To analyze the mechanisms of stabilization of W/W emulsion one should investigate the droplet-droplet interactions – a traditional approach in foam and water-oil based emulsion studies (Exerowa et al. 2018; Tadros 2016; Langevin, 2023). It was shown for W/W emulsions with polymer phases here denoted simply A and B that the importance of the particles contact angle in emulsion stability should be rationalized by its value – particles partitioned in phase A adsorb at droplets from phase B at contact angles $\theta > 90°$ and the A-in-B type of emulsion is more stable than the inverse B-in-A one (Nguyen et al., 2015; Gonzalez-Jordan et al., 2018). The former situation is illustrated in Fig. 10B showing that effective inter-droplet repulsion between the particle adsorption layers on two interacting droplets may prevent coalescence. This scenario seems less likely for B-in-A emulsions, where the particles at $\theta < 90°$ protrude towards the droplet phase (not shown). The considered inter-droplet repulsion originates from the action of steric and/or electrostatic (charged BLG microgels at pH 3) repulsive surface forces (Exerowa et al. 2018; Tadros 2016; Langevin, 2023). However, such inter-droplet repulsion is supposed to effectively act only at sufficient particle surface coverages (Gonzalez-Jordan et al., 2018). Another possible mechanism of stabilization of the A-in-B type of emulsion is inter-droplet bridging by particles (Meng & Nicolai, 2023) as illustrated in Fig. 10C – a phenomenon observed previously for water-oil based emulsion systems (Denkov et al., 1992; Horozov et al., 2005).

The above considerations suggest that the DEX-in-PEG emulsions investigated here at pH 3 are expected to exhibit lower stabilities if the particle contact angles are $\theta < 90°$, which can be achieved in the following cases: 1) pH is set to say pH 7 – a situation where the BLG microgels will partition preferentially into the DEX droplets phase; or 2) keeping pH 3, but producing the inverse emulsion, where the BLG microgels will partition preferentially into the PEG droplets phase. Realization of such experiments with BLG microgel particles of various sizes by the technique used in this work would bring valuable information in direction to generalization the current findings.

## 4. Conclusions

In this study we used a single ATPS to form emulsions with 75 kDa DEX (21.5 % wt.) droplets (volume fraction of 34 % vol.) in 10 kDa PEG (12.5 % wt.) with added BLG particles of different sizes (*ca*. 40–190 nm). The early-stage time evolution of these emulsions was monitored for 75 min. It was found that the process of segregative phase separation evolved through two concurrent processes of droplets coalescence and sedimentation, which may proceed either in parallel or solely. The maximum rates of coalescence-driven and sedimentation-driven phase separation were detected within the time range of about half an hour from the emulsion preparation and were quantified by the maximum velocity $V_{max}$ of the displacement of the relevant advancing fronts. The effect of pH (2–6) on the formation and $V_{max}$ of such emulsions was tested for a single particle size ($d_{h,O} \approx 64$ nm) and the most stable emulsion was obtained at pH 3 (note, the particle size has increased to $d_{h,S} \approx 81$ nm due to swelling) (Movie 1 in Supplementary Materials). Next, experiments with emulsions at pH 3 were performed for 0.1 % wt. BLG particles of different sizes $d_{h,S} \approx$ 40–190 nm (Movie 2 in Supplementary Materials). Supporting results from interfacial tension measurements revealed that the interfacial tension of the protein-free ATPS decreases when BLG particles are present in the system. Efficient droplet coalescence inhibition was found for microgel particles larger than 60 nm, which adsorb on the DEX droplets at contact angles $\theta > 90º$. The possible mechanisms for preventing droplets coalescence were discussed to originate from inter-droplet repulsion between the particle adsorption layers at the droplet surfaces and/or inter-droplet bridging by particles (Gonzalez-Jordan et al., 2018; Meng & Nicolai, 2023).

To generalize the current findings for the particular PEG/DEX ATPS used in this work, further investigations are needed, where formulations from various tie-lines from the PEG/DEX phase diagram (Forciniti et al., 1990, 1991) are employed. In this direction, we performed preliminary experiments with PEG/DEX ATPS at pH 3 using several combinations of polymers with different molecular weights presented in Fig. S3 (Supplementary Materials). The results from interfacial tension measurements for the protein-free ATPS samples proved the variety of tie-lines achieved. At the same time, the results from the phase separation tests with emulsions at pH 3 containing BLG(81) microgel particles revealed weak effects of the polymers $M_w$ on the coalescence-driven phase separation behavior as monitored by the parameter $V_{max}$ of the lower front. In a wider viewpoint, the generalization of this correlation would require examination of various ATPS and particles with different surface chemistries.

**Acknowledgements**

A.B., G.G. and J.Z. gratefully acknowledge the financial support by Polish National Science Centre (Grant No.: Sonata-Bis 2020/38/E/ST8/00173) and (Grant No.: Miniatura 2023/07/X/ST4/01335). D.T. gratefully acknowledges support from the Centre national d'études spatiales (CNES). A.B. gratefully acknowledges support from the Erasmus+ program for funding his scientific visit to the laboratory in Montpellier, where spinning drop experiments were conducted (Grant No.2022-1-PL01-KA131-HED-000057420 and 2023-1-PL01-KA131-HED-000131849).**Author Contributions**
**Andrzej Baliś**: Conceptualization, Data curation, Investigation, Methodology, Visualization, Writing – original draft, Writing – review & editing. **Georgi Gochev:** Conceptualization, Supervision, Visualization, Writing – original draft, Writing – review & editing. **Domenico Truzzolillo**: Data curation, Investigation, Methodology, Visualization, Writing – review & editing. **Dawid Lupa**: Data curation, Investigation, Visualization, Writing – review & editing. **Liliana Szyk-Warszyńska** Data curation, Investigation,

Visualization, Writing – review & editing. **Jan Zawala**: Conceptualization, Funding acquisition, Methodology, Project administration, Resources, Supervision, Writing – review & editing.

**Conflicts of interest**

There are no conflicts to declare

Zhou, T., Liu, Z., Ma, X., Cen, C., Huang, Z., Lu, Y., Kong, T., & Qi, C. (2024). Thermally-resilient, phase-invertible, ultra-stable all-aqueous compartments by pH-modulated protein colloidal particles. *Journal of Colloid and Interface Science*, *665*, 413–421. https://doi.org/10.1016/j.jcis.2024.03.155

# Supplementary Materials

# Water-in-water PEG/DEX/protein microgel emulsions: effect of microgel particle size on the rate of emulsion phase separation

Andrzej Balis,[1] Georgi Gochev,[1,2]* Domenico Truzzolillo,[3] Dawid Lupa,[4] Liliana Szyk-Warszynska,[1] Jan Zawala[1]

[1] *Jerzy Haber Institute of Catalysis and Surface Chemistry, Polish Academy of Sciences, 30-239 Krakow, Poland*

[2] *Institute of Physical Chemistry, Bulgarian Academy of Sciences, 1113 Sofia, Bulgaria*

[3] *Laboratoire Charles Coulomb, UMR 5221, CNRS-Université de Montpellier, F-34095 Montpellier, France*

[4] *Jagiellonian University, Faculty of Physics, Astronomy, and Applied Computer Science, M. Smoluchowski Institute of Physics, Łojasiewicza 11, 30-348 Kraków, Poland*

* Corresponding author: georgi.gochev@ikifp.edu.pl

# Tables

**Table. S1.** DLS experiments: values of the hydrodynamic diameters $d_h$ of BLG aggregates after heating at 85°C in a water bath for 20 minutes without stirring.

| Sample: pH of initial native BLG solutions before heating | Z-average $d_h$ [nm] | Standard Deviation [nm] | PDI [%] | Mean size by intensity (peak 1. 100% area) [nm] | Standard Deviation [nm] | Mean size by number (peak 1. 100% area) [nm] | Standard Deviation [nm] |
|---|---|---|---|---|---|---|---|
| | *$d_{h,O}$ | | | | | | |
| 5.7 | **713** | **9.3** | 33 | 665 | 47.7 | 618 | 44.6 |
| 5.8 | **162** | **3.3** | 7 | 174 | 4.2 | 135 | 2.9 |
| 5.9 | **101** | **1.3** | 17 | 109 | 5.2 | 65 | 6.1 |
| 6.0 | **64** | **0.3** | 8 | 68 | 0.2 | 50 | 1.0 |
| 6.1 | **60** | **0.7** | 7 | 65 | 1.9 | 45 | 2.2 |
| 6.2 | **51** | **1** | 10 | 56 | 0.2 | 36 | 0.5 |
| 6.3 | **48** | **0.7** | 8 | 52 | 0.3 | 35 | 1.4 |
| 6.4 | **28** | **0.3** | 23 | 33 | 0.9 | 17 | 0.6 |
| | ** $d_{h,S}$ | | | | | | |
| 5.7 | **1077** | **260** [a] | 75 | 417 | 113.8 | 409 | 107.4 |
| 5.8 | **191** | **3** | 6 | 205 | 2.1 | 165 | 9.2 |
| 5.9 | **123** | **0.8** | 16 | 125 | 7.1 | 90 | 5 |
| 6.0 | **81** | **0.5** | 6 | 86 | 0.7 | 62 | 2.3 |
| 6.1 | **78** | **0.3** | 6 | 84 | 1.1 | 58 | 0.3 |
| 6.2 | **69** | **1.1** | 10 | 75 | 1.2 | 49 | 3.3 |
| 6.3 | **61** | **1.5** | 16 | 62 | 2.0 | 44 | 2.4 |
| 6.4 | **42** | **1.8** | 23 | 42 | 2.2 | 28 | 0.4 |

\* Original BLG aggregates dispersions;. hydrodynamic diameters denoted $d_{h,O}$
\*\* BLG aggregates dispersions set to pH 3; hydrodynamic diameters denoted $d_{h,S}$
[a] Anomalously high value

**Figures**

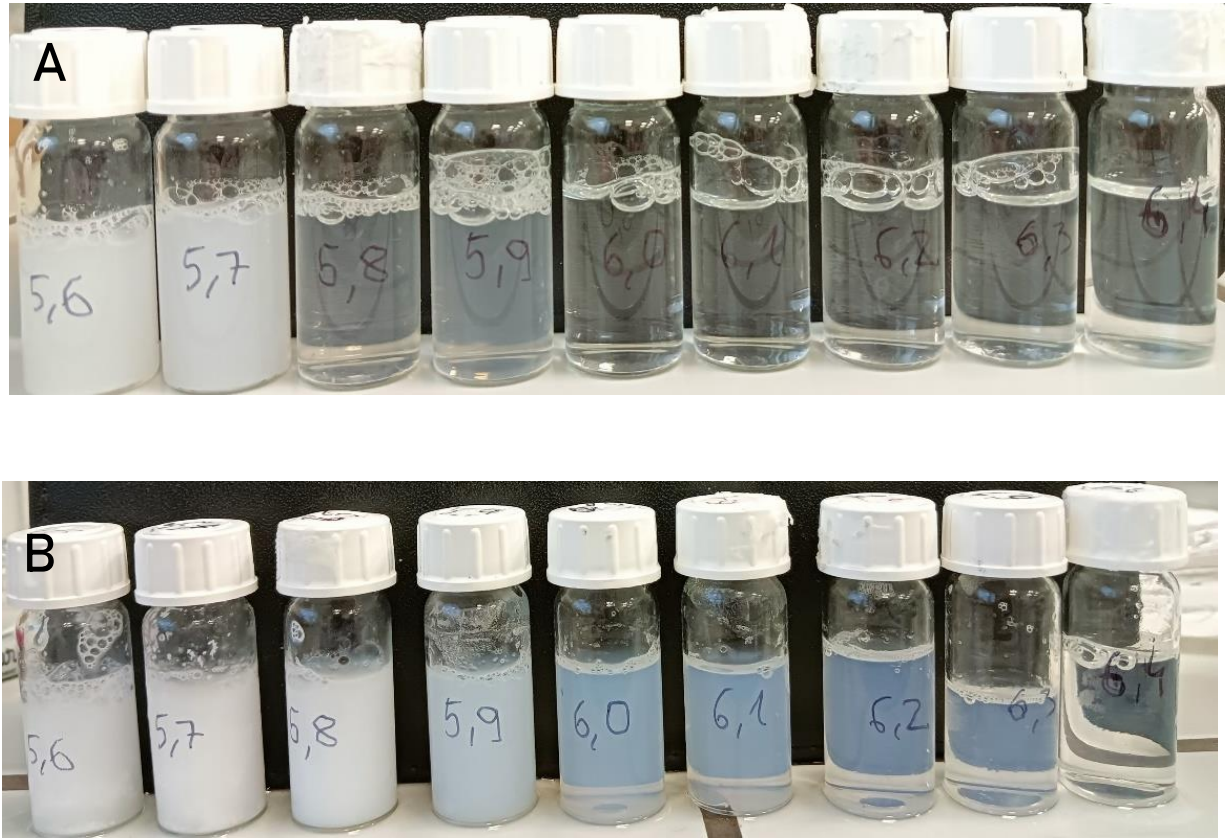

**Fig. S1. A.** BLG native protein solutions adjusted to pH values indicated on the bottles. **B.** The respective suspensions of BLG aggregates after heating at 85°C in a water bath for 20 minutes without stirring.

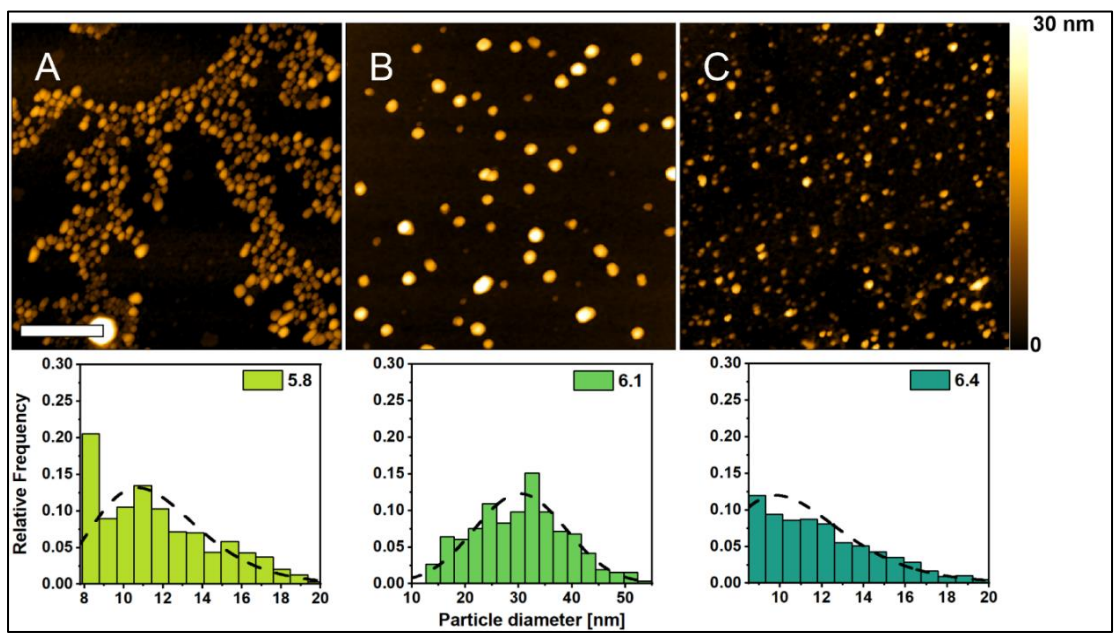

**Fig. S2.** (top panel) Representative topographical AFM images of BLG-gels obtained under different values of initial suspension and deposited at mica: pH 5.8 (A), 6.1 (B) and 6.4 (C). Deposition conditions: 0.001 M HCl (pH 3). The size distribution corresponding to the initial pH value is given below each image. Dashed lines denote the distribution function fitted to experimental data: log-normal (pH 5.8 and 6.4) and normal (pH 6.1). Each distribution was derived based on the analysis of at least 1000 particles. The scale bar corresponds to 500 nm. The false color scale presented at the right side is applicable to all the presented images. (bottom panel) Number-weighted size distribution of BLG nanoparticles prepared under different conditions of initial pH value.

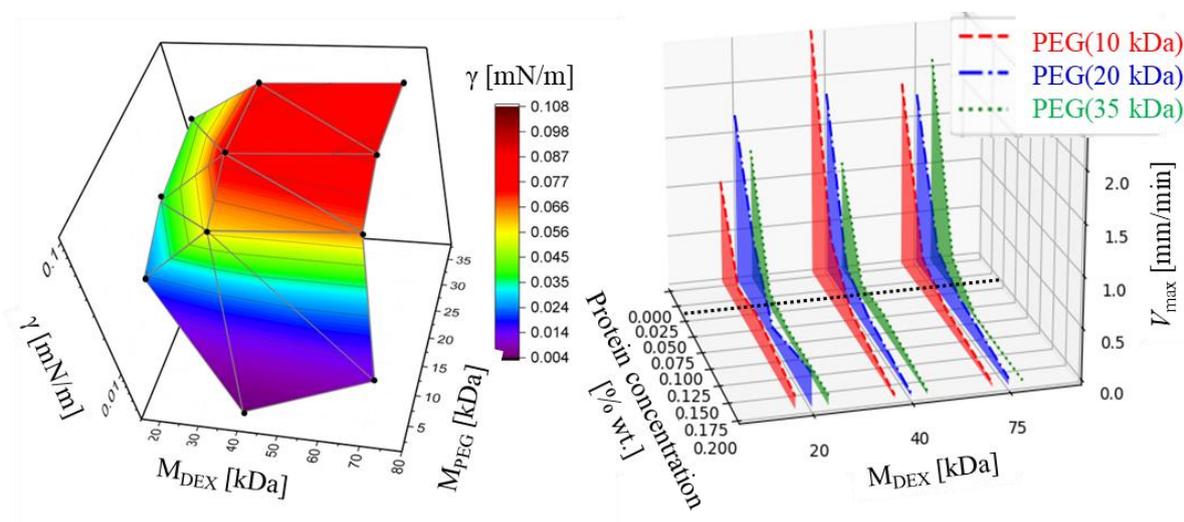

**Fig. S3.** (left) 3D-plot showing the dependence of the interfacial tension of PEG/DEX ATPS on molecular weight of PEG ($M_{PEG}$) and DEX ($M_{DEX}$) polymers. (right) The dependence of the maximum velocity $V_{max}$ of displacement of the lower front in DEX-in-PEG emulsions on the molecular weight of polymers and the concentration of microgel particles BLG(81), evaluated from the $h(t)$ profiles shown below.

# DEX-in-PEG emulsions

DxPy ≡ DEX(x kDa)PEG(y kDa); cBLG is the concentration of microgel particles BLG(81)

## DEX(20 kDa) serries

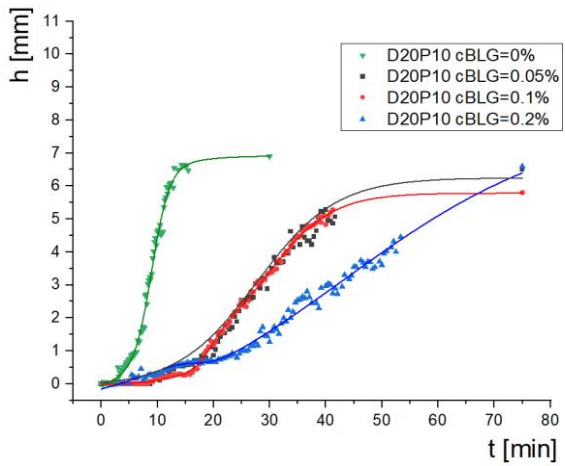
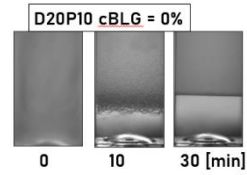
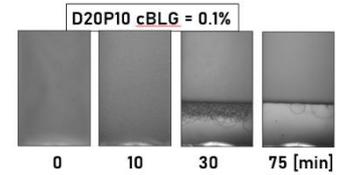
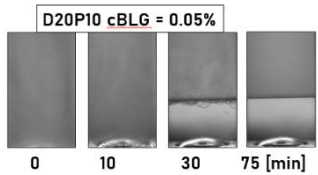
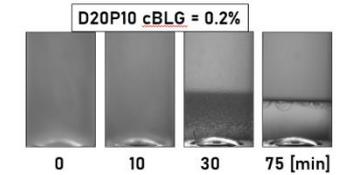

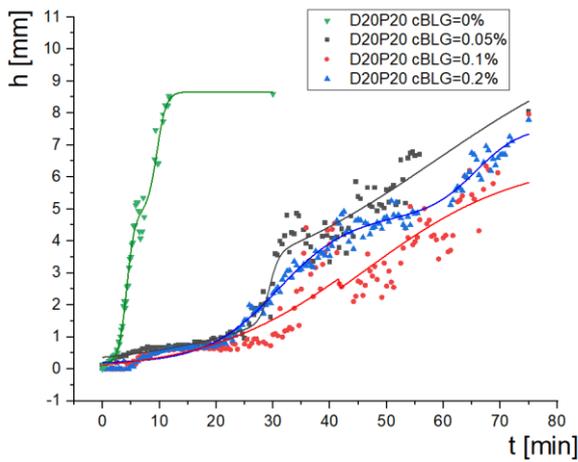
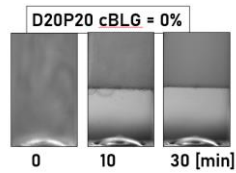
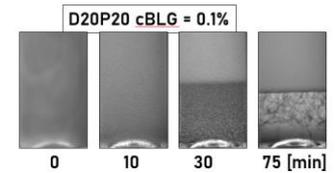
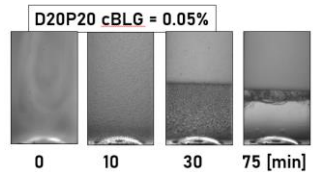
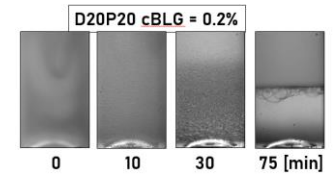

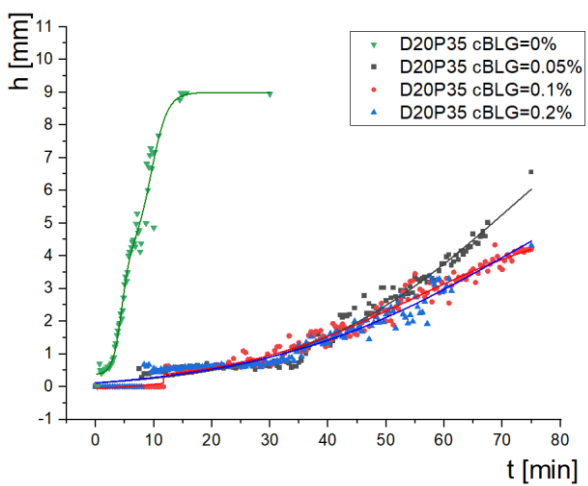
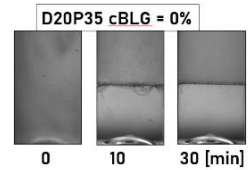
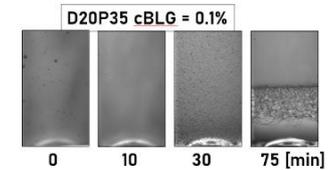
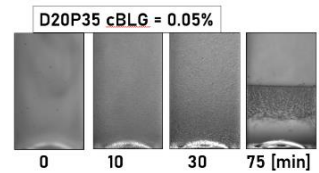
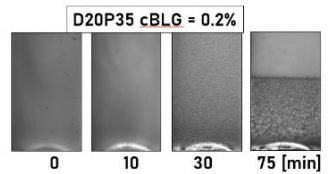

## DEX(40 kDa) serries

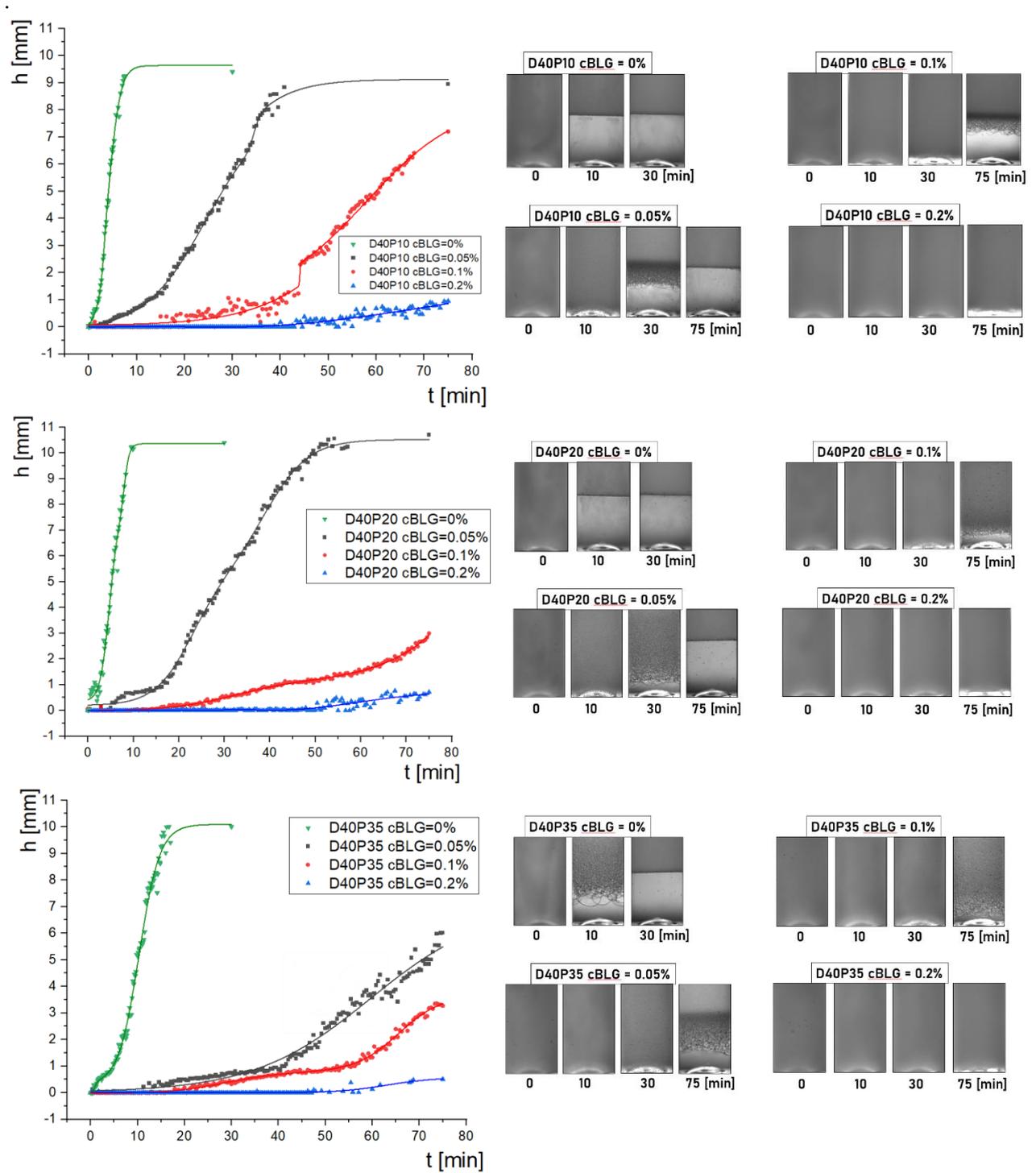

## DEX(75 kDa) serries

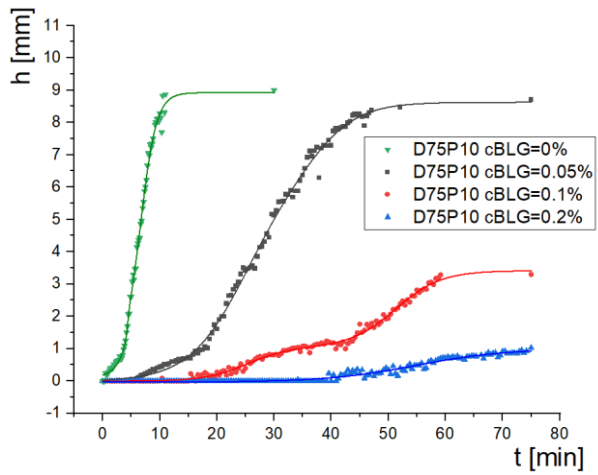
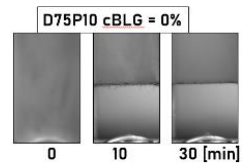
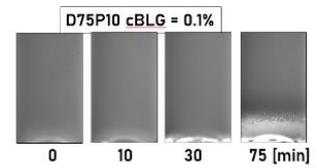
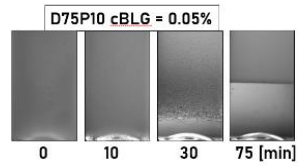
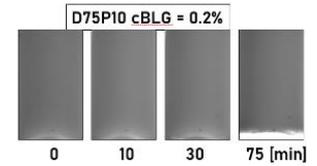

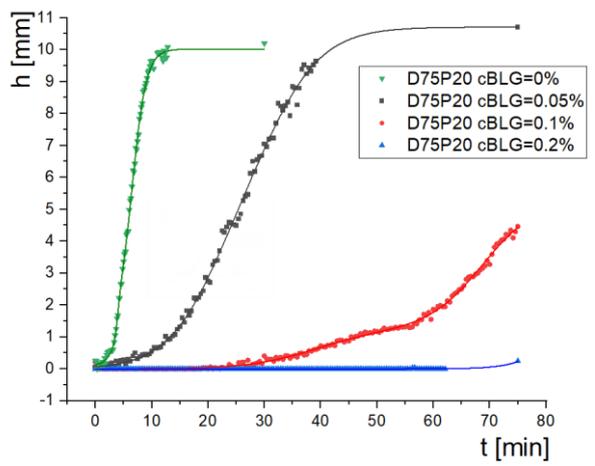
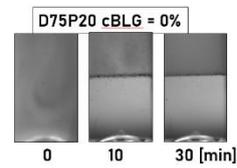
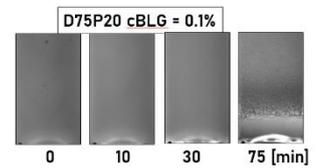
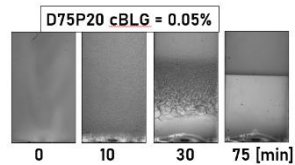
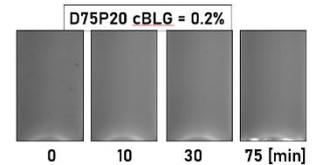

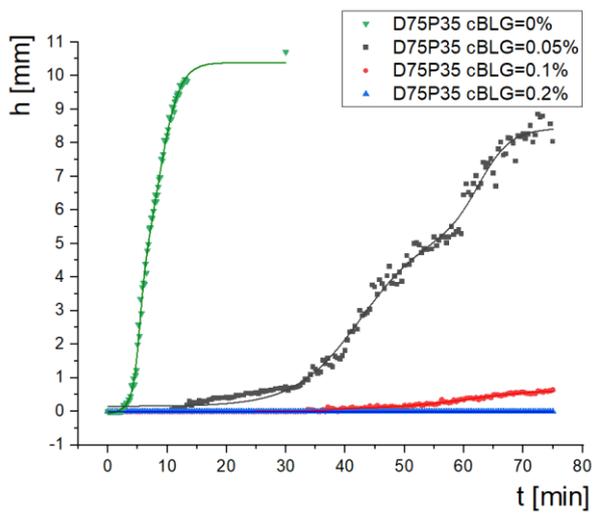
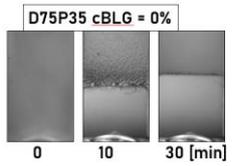
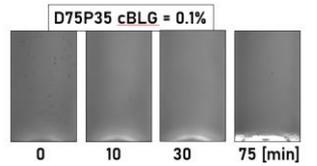
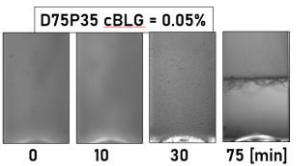
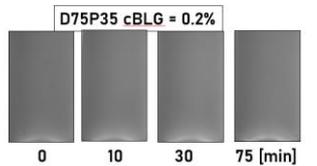